\documentclass{PoS}

\title{Insights on the physics of SNIa obtained from their gamma-ray emission}

\ShortTitle{Gamma Rays from SNIa}

\author{\speaker{J. Isern} \\
        Institut de Ci\`encies de l'Espai (ICE-CSIC) \& Institut d'Estudis Espacials de Catalunya (IEEC) \\
        E-mail: \email{isern@ice.cat}}

\author{E. Bravo\\
        ETSAV, Universitat Polit\`ecnica de Catalunya \\
        E-mail: \email{eduardo.bravo@upc.edu}}

\author{P. Jean, J. Knodlseder\\
         IRAP \& Universit\'e de Toulouse\\
       E-mail: \email{pjean@irap.omp.eu, Jurgen.Knodlseder@irap.omp.eu}}

\abstract{Type Ia supernovae are thought to be the outcome of the thermonuclear explosion of a carbon/oxygen white dwarf in a close binary system. Their optical light curve is powered by thermalized gamma-rays produced by the radioactive decay of $^{56}$Ni, the most abundant isotope present in the debris. Gamma-rays escaping the ejecta can be used as a diagnostic tool for studying the structure of the exploding star and the characteristics of the explosion. The fluxes of the $^{56}$Ni lines and the continuum obtained by INTEGRAL from 
SN2014J in M82, the first ever gamma-detected SNIa, around the time of the 
maximum of the optical light curve strongly suggest the presence of a plume of $^{56}$Ni in the outermost layers moving at high velocities. If this interpretation was correct, it could have important consequences on our current understanding of the physics of the explosion and on the nature of the systems that explode.
}

\FullConference{11th INTEGRAL Conference Gamma-Ray Astrophysics in Multi-Wavelength Perspective,\\
		10-14 October 2016\\
		Amsterdam, The Netherlands}

\begin{document}

\section{Introduction}
There is a wide consensus that Type Ia supernovae (SNIa) are caused by the explosion of a carbon-oxygen white dwarf \cite{hoyl60} near the Chandrasekhar's mass in a close binary system. This idea is supported by the absence of hydrogen in the spectra, the presence of SNIa in all kind of galaxies, including elliptical, where star formation stopped long time ago, the compact character of the exploding object imposed by the rapid rise of the light curve, and the noticeable spectrophotometric homogeneity of the events. These supernovae are called \emph{Branch-normal} and their main properties are the clear correlation between the brightness at maximum and the post-peak decline rate of the light curve \cite{psko77,psko84,bran82,phil93}, and the existence  of a secondary peak in the near infrared light curve $\sim$20 days after the optical maximum \cite{meik00}. A careful examination of the SNIa events have revealed that only  $\sim 70$\% have this normal behavior and that the remaining ones display large deviations  from it \cite{li11a,li11b}. Consequently, three additional subtypes added: i) SN1991T-like. They display light curves with very bright and broad peaks and  they are very often associated to interactions with the circumstellar material. This class represents the $\sim 9$\% of all SNIa events. ii) SN1991bg-like. They are characterized by dim and narrow optical peaks, as well as by the absence of the secondary peak in the infrared. This group contains $\sim 15$\% of the events. iii) SN2002cx-like or Type Iax. They exhibit a high-ionization level during the pre-maximum but have luminosities similar to those of 91bg-like, velocities that are roughly half of the normal ones, and do not display a secondary maximum in the infrared. iv) Unusual cases like Ca-rich transients, fast-declining transients, slowly declining events like '02-es' or super-Chandrasekhar events \cite{taub17}. Despite the strange behavior of some of these events, the resulting nucleosynthesis and the nature of the hosting galaxies strongly suggest they have a thermonuclear origin. Thus, different scenarios and burning mechanisms could be necessary to account for these explosions.

From the point of view of the explosion mechanism, and under the spherical symmetry hypothesis, it is possible to distinguish four cases. Pure detonation models in which the flame propagates supersonically, pure deflagration models in which the flame propagates subsonically, delayed detonation models in which the flame initially propagates as a deflagration and, when the density falls below a critical value, it turns out into a detonation, and pulsating detonation models in which the flame initially propagates so slowly that the star expands and induces the quenching of the flame and later on the detonation of the unburnt fuel when the star recontracts. In 3D there are also similar behaviors but with a wider variety of possibilities. 

Several scenarios have been advanced up to now: i) In the single degenerate scenario (SD) the white dwarf accretes matter from a non-degenerate companion  and explodes when it reaches the critical mass; the accreted matter can be either hydrogen or helium \cite{whel73,nomo82,han04}. ii) in the double degenerate scenario (DD) two white dwarfs merge as a consequence of the momentum losses caused by the emission of gravitational waves; the evolution of the merger is not completely understood at present and consequently it is not known at which moment the explosion will occur \cite{webb84,iben84}. iii) In the sub-Chandrasekhar scenario (SCH) it is assumed  that a C/O white dwarf, with a mass not necessarily near the critical one, accretes helium and detonates as a consequence of the shock wave generated by the ignition of the bottom of the freshly accreted layer \cite{woos94,livn95}; this helium can be directly accreted from a non-degenerate He-star or He-white dwarf, or it can accumulate in the outer layers as the product of the burning of the hydrogen that is being accreted. iv) In the white dwarf-white dwarf collision scenario (WD-WD) it is assumed that two white dwarfs collide and immediately ignite 
\cite{pakm12,kush13}. v) In the core degenerate scenario (CD) the white dwarf merges with the core of an AGB star; this case corresponds to the prompt merger in the DD scenario, and the explosion can occur at any time after the merger \cite{livi03, kash11}. 

During the explosion, important amounts of radioactive isotopes are produced, the must abundant being $^{56}$Ni. The thermalization of the gamma-ray photons and the annihilation of the positrons produced during the decay chain $^{56} {\rm Ni} \rightarrow ^{56}{\rm Co} \rightarrow ^{56}{\rm Fe}$ provide the major part of the energy that powers the light curve of those SNIa that are not interacting with the interstellar medium as it was proposed by \cite{pank62,colg69} and confirmed by \cite{chur14}. As ejecta expands, matter becomes more and more transparent and an increasing number of gamma-rays escape thermalization and can be used as a diagnostic tool for studying the structure of the exploding star and the characteristics of the explosion  \cite{clay69,gome98,hofli98,the14}, since the amount and distribution of the radioactive material strongly depend on how the ignition starts and how the nuclear flame propagates \cite{gome98,iser08}. The advantage of using $\gamma$-rays  for diagnostic purposes comes from their ability to distinguish among different isotopes and on the relative simplicity of their transport modelling. Notice, however, that in the case of close enough supernova outbursts, less than $\sim 1$~Mpc, it is possible to obtain high-quality $\gamma$-ray spectra, and to perform detailed comparisons with theoretical predictions. However, when realistic distances are taken into account only some outstanding features, like the intensity of the lines or of the continuum, have the chance to be detected \cite{gome98}. 

Several authors have examined the $\gamma$-ray emission from SNIa predicted by different models. To explore the model variants we used the code described in \cite{gome98}, which was successfully cross-checked with the results obtained by other authors \cite{miln04}. Before and around the maximum of the optical light curve, the $\gamma$-emission can be characterized  as it follows: i) a spectrum dominated by the $^{56}$Ni 158 and 812 keV lines, ii) because of the rapid expansion, the lines are blueshifted, but their energy peak quickly evolves back to the red as matter becomes more and more transparent; the emergent lines are broad, typically from 3\% to 5\%, iii) the intensity of the $^{56}$Ni lines rises and declines very quickly with time as a consequence of the rapid expansion of the debris and its relatively short decay time (T$_{1/2}\sim 6$ days), and are very weak at the beginning even in the SCH models. Because of the Doppler effect, the 812 keV line blends with the $^{56}$Co 847 keV line and forms a broad feature that declines more slowly.

In all these spherically symmetric models, $^{56}$Ni is buried in the inner layers and it is necessary to wait for a substantial expansion of the debris to allow the escape of non-thermalized $\gamma$-photons. Despite having important amounts of  $^{56}$Ni in the outer layers, SCH models have a similar behavior. It is important to realise here, that the viability of these SCH models has been questioned as a consequence of the severe constraints posed by the existing optical observations on the total amount of $^{56}$Ni that can be synthesized in these outer layers.

\section{Gamma-ray emission from SN2014J}

\begin{figure}[h]
\begin{minipage}{17pc}
\includegraphics[width=19pc,angle=90]{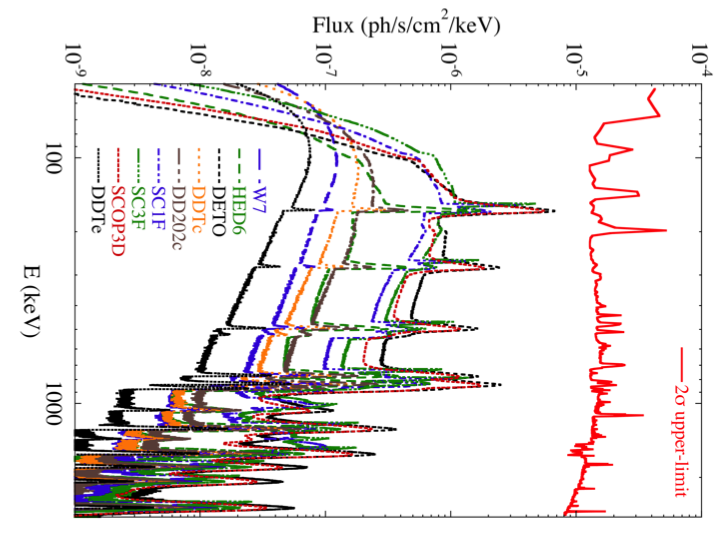}
\caption{\label{fignoise} \footnotesize Comparison between the $2\sigma$ upper limit flux obtained by \emph{SPI/INTEGRAL} around the maximum of the optical light of SN2011fe with the mean flux predicted by several 1D models at the same epoch \cite{iser13}.}
\end{minipage}\hspace{2pc}%
\begin{minipage}{17pc}
\includegraphics[width=17pc]{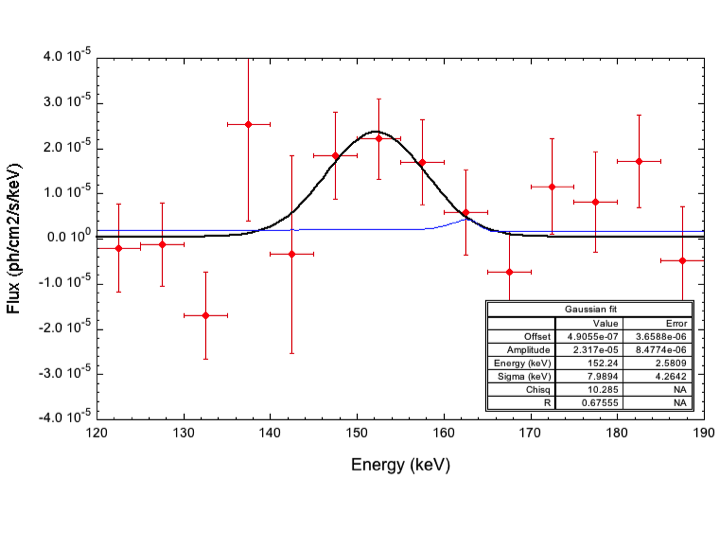}
\caption{\label{figspi} \footnotesize Spectrum of the SN2014J obtained by \emph{SPI} during revolutions 1380-81 in bins of 5 keV in the 120-190 kev energy band. The black line represents the gaussian fit of the data.}
\end{minipage} 
\end{figure}

As a consequence of the poor sensitivity of the instruments to the broad lines and the scarcity of bright events, we had to wait for SN2014J, the brightest Type Ia supernova since the Kepler event, to detect the gamma-emission of SNIa \cite{chur14a,dieh14,iser14}. In between we have only been able to obtain upper limits to the gamma-ray emission of these supernovae. These are the cases of SN1991T \cite{lich94} and SN1998bu \cite{geor02}, observed with \emph{COMPTEL}, and SN2011fe observed with \emph{INTEGRAL} \cite{iser13}. As en example, Figure~\ref{fignoise} displays the flux obtained by \emph{SPI/INTEGRAL} in the case of SN2011fe compared with the signal predicted by different 1D models \cite{iser13}.

The behaviour of the instrumental background produced by the interaction of the cosmic rays  and solar protons with the instrument is complex but, in principle, the background lines can be distinguished from the signal thanks to the spatial and temporal modulation produced by the coded mask and dithering. Nevertheless, despite precautions,  some instrumental residual lines could remain if the background is not correctly modelled. Therefore, since these lines are intrinsically narrow, any narrow feature present in the observed spectrum is suspicious of having an instrumental origin. This is the case of the two main decay lines of  $^{56}$Ni, the 158 and 812 keV lines, which can be confused with two \emph{SPI} instrumental lines caused by the decay of $^{47}$Sc and $^{48}$Co that produce lines at 159 keV and 811 keV respectively (see \cite{iser16}). 

The second difficulty that have to face the observations with both instruments, \emph{IBIS/ISGRI} and \emph{SPI}, is that he flux extracted in an energy bin contains not only the corresponding photons emitted by the source at this energy, but also events produced by photons of higher energy that do not deposit all their energy in the detector (e.g. by Compton scattering). This last contribution is not negligible at low energies and, in order to compute its value, it is necessary to convolve the response of the instrument with the expected theoretical spectra \cite{iser16}.

Finally, the third difficulty comes from the temporal variability of the $^{56}$Ni lines that does not allow to integrate the signal for a long time interval. In the limit of weak signals, the optimal integration time is estimated to be $\Delta t \sim 1.26\tau$, where $\tau$ is the characteristic growing/declining time of the line. For instance, in the case of 1D delayed detonation models this time is of the order of a week \cite{iser13}. 

\emph{IBIS/ISGRI} covers a roughly similar energy range as \emph{SPI} but its efficiency begins to drop above 100 keV and at 812 keV its sensitivity is much worse than \emph{SPI}. Below 100 keV the sensitivity is of the same order or even larger. As mentioned before, the contribution of the secondary photons is important and since the spectra of supernova is not equivalent to that of the Crab, the usual procedure of normalization to the Crab values is no longer valid and a fully convolution with theoretical spectra has to be performed to obtain the correct values \cite{iser16}.  

SN2014J was discovered by \cite{foss14} on January 21st 2014 in M82 ($d=3.5\pm 0.3$)~Mpc. Three observation runs with \emph{INTEGRAL} were performed. The first one started 16.5 days after the explosion (a.e., from now) and finished 35.2 days a.e, the second one covered the period 50-100 days a.e., when the spectrum of the supernova is dominated by the cobalt lines, and the third one the period 130-162 a.e.


The analysis of the data obtained by SPI during this first observation period revealed an emission excess, with a significance $\sim 5\,\sigma$, in the 70-190 and the 650-1300 keV bands at the position of of SN2014J that was not present before the explosion and that was clearly isolated from the surrounding sources. In the energy band around 158 keV $^{56}$Ni, i.e, 120 - 190 keV, a broad and completely unexpected redshifted feature\footnote{See \cite{dieh14} for a different interpretation of the data}  was detected (Fig.~\ref{figspi}). This feature was characterized by a flux $(1.6 \pm 0.4) \times 10^{-4}$  ph s$^{-1}$ cm$^{-2}$  centered at $155.2 ^{+1.3}_{-1.1}$~keV and with a FWHM of $5.2 ^{+3.4}_{-2.2}$~keV.   

If the data are grouped into bins corresponding to the revolutions 1380-81 (16.5 - 22.2 days a.e.), 1382-83 (22.6 - 28.2 days a.e.), and  1384-85 (28.6 - 34.2 days a.e.), the Gaussian fit gives $(2.23 \pm 0.8) \times 10 ^{-4}$ ph s$^{-1}$ cm$^{-2}$ centered at $152.6 \pm 2.8$~keV and a significance of 2.8 $\sigma$ for the first bin, and only 2 $\sigma$ upper limits of  $< 1.72 \times 10 ^{-4}$ and  $<2.23 \times 10 ^{-4}$ ph s$^{-1}$ cm$^{-2}$ for the other two bins respectively. If the complete response of SPI is adopted, the flux in the line becomes $(1.59 \pm 0.57) \times 10 ^{-4}$, centered at $154.5 \pm 0.64$ keV and a width of $ 3.7 \pm 1.5$~eV for the first bin, and $< 1.42 \times 10 ^{-4}$ and  $<1.52 \times 10 ^{-4}$ ph s$^{-1}$ cm$^{-2}$ for the remaining ones.

\begin{table}
\centering
\begin{tabular}{ c c c c}
   \\ \hline \hline
    Revolutions & Days & counts (s$^{-1}$) & S/N
\\ \hline  
1380-81 & 22.20 -16.50 & $0.149\pm 0.039$ & 3.8 \\
1382-83 & 28.20 -22.60 & $0.078\pm 0.041$ & 1.9 \\
1384-85 & 34.20 - 28.60& $0.143\pm 0.037$ & 3.8 \\
\\ \hline
   \end{tabular}
    \caption{Temporal evolution of the 158 keV $^{56}$Ni according to IBIS/ISGRI}
    \label{tabibis}
\end{table}

The data obtained by \emph{IBIS/ISGRI} suggest a similar pattern. There is a 3.8 $\sigma$ emission excess during orbits 1380-81 and 1384-85, separated by a dip that is compatible with the free decay of $^{56}$Ni (see table~\ref{tabibis}). This behavior is similar to that obtained by \emph{SPI} but has a better significance to the point that the upturn at the end of the exposure seems real, suggesting that new radioactive layers were exposed. However, the poor S/N of the central bin prevents any solid conclusion about this point and an approximately constant or gently decaying behavior cannot be excluded \cite{iser16}.

\begin{figure}[h]
\begin{minipage}{17pc}
\includegraphics[width=17pc]{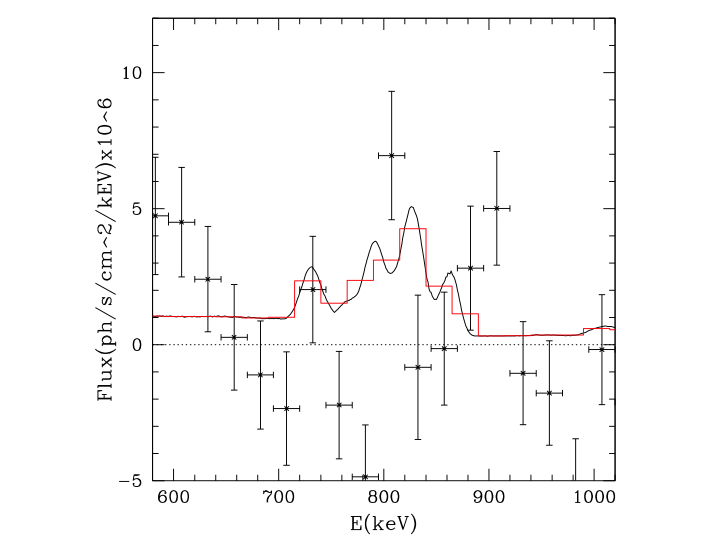}
\caption{\label{fig6b} \footnotesize Spectrum of the SN2014J obtained by \emph{SPI} during revolutions 1380-81 in bins of 25 keV. The black line represents the expected emission from the 3Dbball model for the same epoch. The red line is the same but in bins of 25 keV}
\end{minipage}\hspace{2pc}
\begin{minipage}{17pc}
\includegraphics[width=17pc]{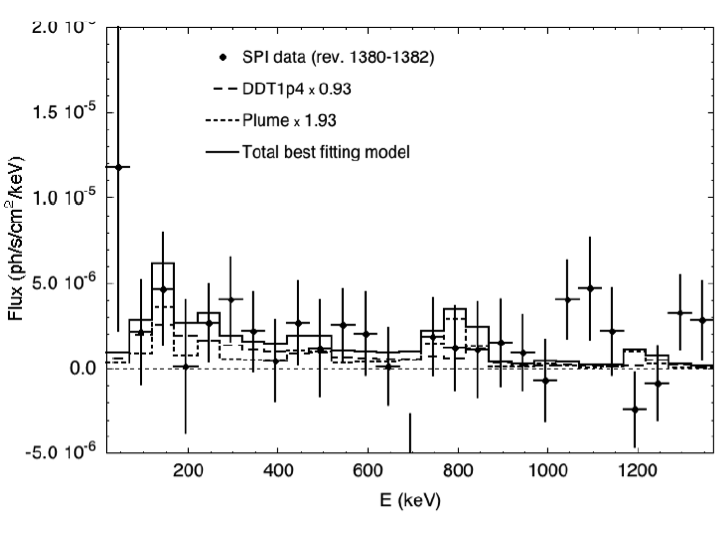}
\caption{\label{figbb} \footnotesize  Gamma ray spectrum during revolutions 1380-82 (16.5-25.0 days a.e.).Bins are 50 keV wide.The continuous line represents the best fit obtained scaling the model DDT1p4 by a factor 0.93 (0.605 M$_\odot$ of $^{56}$Ni -long dashed line- and adding a $^{56}$Ni plume of 0.077 M$_\odot$ -short dashed line.}
\end{minipage} 
\end{figure}

There is also a flux excess in the 720-870 keV band. This excess has a significance of $\sim\,2.8\sigma$ and can be attributed to the contribution of the $^{56}$Ni-$^{56}$Co decays. The blending of the 812 keV $^{56}$Ni and the 847 keV $^{56}$Co caused by the Doppler broadening \cite{gome98} together with the relative weakness of the fluxes prevents any spectroscopic analysis of the individual $\gamma$-lines. Here it is important to notice the presence of a $\sim 2.6\, \sigma$ feature at $\sim 730$ keV, that is the expected position of the 750 keV $^{56}$Ni if it has the same redshift as the 158 keV line (Fig.~\ref{fig6b}). The Gaussian fit of this feature gives a flux of $(1.57 \pm0.7) \times 10 ^{-4}$ ph s$^{-1}$ cm$^{-2}$ ($ 2.1 \sigma$), a centroid placed at $733.4 \pm 3.8$ keV, and a FWHM of $16.9 \pm 9.0$ keV.  

During the second run of observations, the 847 and 1238 keV $^{56}$Co emission lines were detected for the first time in a SNIa \cite{chur14}. The observed fluxes were $(2.34\pm 0.74)\times 10^{-4}$ and $(2.78\pm 0.74)\times 10^{-4}$ ph~cm$^{-1}$s$^{-1}$, respectively.  The lines where placed at $852\pm 4.5$ and $1255 \pm 7$ keV and the FWHM was $24 \pm 8$ and $45 \pm 14$ keV, respectively, which represent a broadness of $\sim 3-4\%$
\cite{chur15}.


\section{Discussion and conclusions}

There are several spherically symmetric models, corresponding either to a deflagration or a delayed detonation, that eject a mass of the order of the Chandrasekhar's mass and have the bulk of radioactive elements in the central region of the expanding debris that are broadly consistent with the observed emission of $^{56}$Co in SN2014J at late times (50 days a.e.). For instance, a DDT model -- a model that starts as a central deflagration and makes a transition to a supersonic regime when the density ahead the flame  is low enough -- like the DDT1p4, which produces 0.65 M$_\odot$ of $^{56}$Ni, ejects 1.37 M$_\odot$ and has a total kinetic energy of $1.32 \times 10^{51}$~ergs, predicts an spectrum that broadly agree with the one obtained by \emph{SPI} during the late epoch of observations  and is able to reproduce the observed optical light curve. Pure detonations or 'strong' sub-Chandrasekhar models can be excluded because they overproduce $\gamma$-rays. However, all the models are consistent with the present late data,  if they are normalized just changing the amount of $^{56}$Ni  synthesized and keeping all the other parameters fixed \cite{chur14,chur15}.

In the late spectra there are not evidences of the existence of velocity substructures. In other words, the data are not consistent with the presence of important amounts of $^{56}$Co moving with a small velocity dispersion. The maximum amount of radioactive material that can be present within a bin of $\sim 3$~keV near the 847 keV line is $\lesssim 0.02$~M$_\odot$ at a $2\, \sigma$ level.


Despite the reasonable agreement with the observations, none of these 1D models is able to reproduce the intensity and the redshifted nature of the 158 keV $^{56}$Ni line observed by \emph{SPI} nor the emission excess and temporal behavior found in the \emph{IBIS/ISGRI} and \emph{SPI} data (Fig.~\ref{figbb}, dashed line). Therefore, given the transient behavior of these 158 \& 812 keV Ni lines just before the maximum it seems natural to wonder about  the presence of this isotope in the outermost layers of the expanding debris. Since $^{56}$Ni was not visible in the optical spectrum, this material could not be in the form of a single blob but distributed in 'tiny' fragments in order to avoid the thermalization of the gamma-rays and the subsequent heating and optical emission \cite{iser16}.

\begin{figure}[h]
\begin{minipage}{17pc}
\includegraphics[width=17pc]{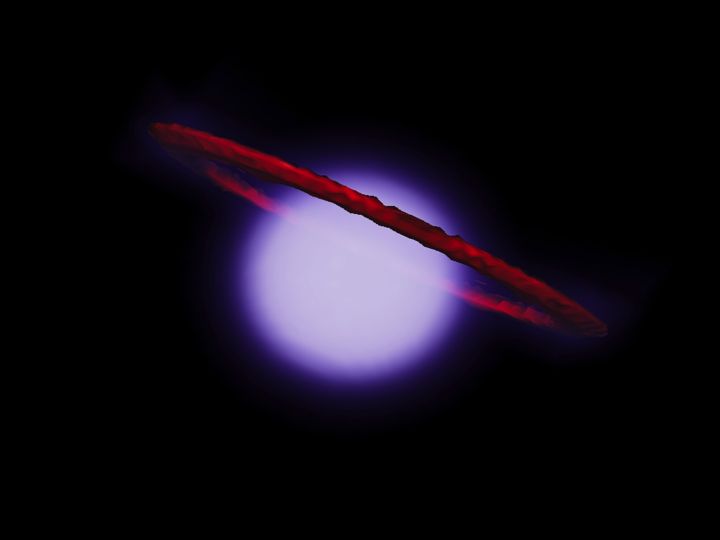}
\caption{\label{fig44} \footnotesize Phenomenological scenario to account for the early $\gamma$-ray emission. It assumes the existence of a central, almost spherical, remnant that contains the bulk of the ejecta and of the radioactive material, plus a conically shaped ring made of almost pure $^{56}$Ni expanding with velocities high enough to avoid being caught by the inner material.}
\end{minipage}\hspace{2pc}%
\begin{minipage}{17pc}
\includegraphics[width=17pc]{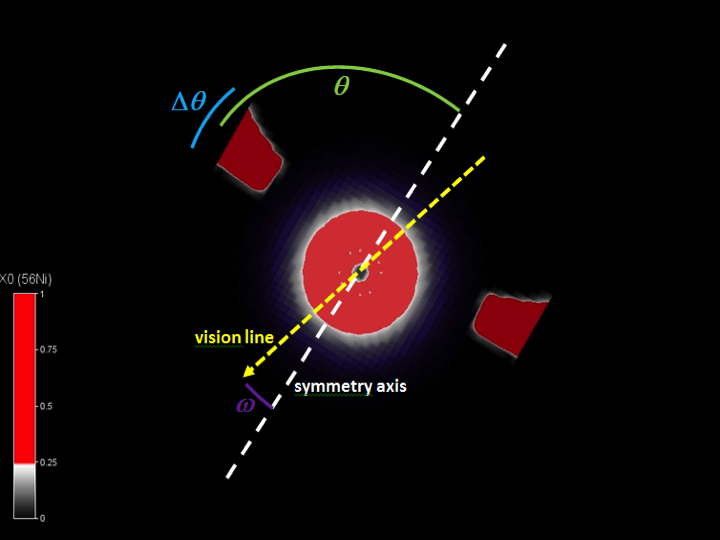}
\caption{\label{figbdm} \footnotesize Geometrical structure of the model displayed in the previous figure.}
\end{minipage} 
\end{figure}

These arguments, together with the constraints imposed by the non-detection of $^{56}$Ni in the optical spectrum at maximum light led to propose a model with a surrounding radioactive ring containing $\lesssim 0.08$~M$_\odot$ of $^{56}$Ni (Fig.~\ref{fig44}). The observed redshift of 3.2 keV suggests that the radioactive material is receding from the observer with an average velocity $v \approx 6,000$ km/s and it is placed in the far hemisphere  while the observed average width, 4.9 keV, of the line indicates a maximum velocity dispersion of 10,000 km/s  along the line of sight. Nothing can be said about the radial velocity of the blobs except that it has to be larger than $\sim 30,000$ km/s in order to not to be caught by the outer layers of the supernova. Several geometries were considered (Fig.~\ref{figbdm}). One, not necessarily the unique, was obtained assuming $\theta \approx 78^o$ and $\Delta \theta \approx 12^o$. The spectrum from the ring (dotted line) and total (continuous line) are displayed in Fig~\ref{figbb}.

The plausibility of this hypothesis is supported by the observed rapid raise of the optical light curve  at early times \cite{zhen14,goob14} and by the microvariability found just after maximum \cite{bona16}. Additional arguments will come from the chemical inhomogeneity found in Kepler \cite{reyn07} and Tycho \cite{vanc95, tseb16} remnants and by the high velocity features detected in the early optical spectra of many SNIa around 10 days a.e. \cite{tana06}.


\begin{figure}
\begin{center}
\includegraphics[width=0.8\textwidth]{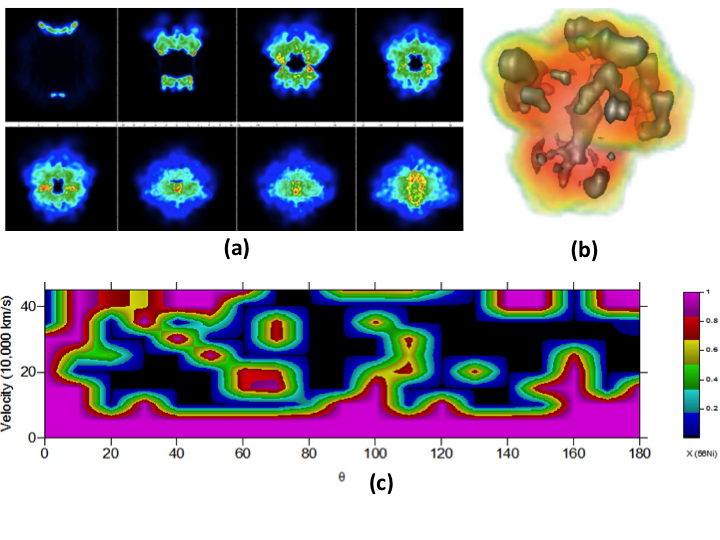}
\caption{\footnotesize a): Asynchronous ignition (0.14 s) of two poles of a He-layer surrounding a C/O white dwarf and the later evolution of the flame until the complete destruction of the star. b): Distribution with the form of a fragmented radioactive ring 100 s after ignition. c): Distribution of $^{56}$Ni as a function of the velocity module and the angle $\theta$ of Fig.~6. }\label{figm}
\end{center}
\end{figure}

The model presented in Fig~\ref{fig44} has been constructed 'ad hoc' to provide an explanation to the observed early emission excess. Certainly, it can look a bit tricky at a first glance but this kind of structures appear in a natural way when we consider asynchronous ignitions of He-mantles surrounding C/O white dwarfs. Fig.~\ref{figm}, panel a), represents the evolution of the temperature when a He-layer on the top of a massive C/O white dwarf is asynchronously ignited in two opposite points. Panel b) shows how a ring structure made of radioactive fragments forms, and panel c) shows the distribution of $^{56}$Ni according the direction of the visual and the velocity module. These fragments, depending on their velocity and on the observer will produce a rich substructure in the spectrum (Bravo et al., in preparation) that could be detected by a sensitive enough detector and/or for a close enough supernova. Notice, however, that since these spectral features depend on the line of sight, it would be necessary to observe them in a big enough sample of supernovae to reach a solid conclusion. In other words, a sensitive enough gamma -ray detector is necessary to capture the information contained in the early spectrum. 

It is clear that, if such excesses are present in other supernovae, models in which the flame starts in the central regions and the radioactive elements are confined in the deep interior are not appropriate and other possibilities have to be explored. As mentioned before, one possibility is provided by the sub-Chandrasekhar models if there is the possibility to trigger the explosion with a small amount of helium in the outer layers \cite{fink10}. Other possibilities are provided by other three dimensional scenarios like pulsating reverse detonations \cite{brav09,brav09b}, gravitationally confined explosions \cite{kase05} or white dwarf collisions in multiple systems \cite{pakm12,kush13,azna13}.

\section*{Acknowledgements}

This  work was  supported by the  MINECO-FEDER grants ESP2013-47637-P \& ESP2015-66134-R (JI), AYA2013-40545 (EB), by the grant 2009SGR315 and the CERCA program of the Generalitat de Catalunya (JI).
The SPI project has been completed under the responsibility and leadership of CNES, France. ISGRI has been realized by CEA with the support of CNES.
We acknowledge the \emph{INTEGRAL} Project Scientist Erik Kuulkers (ESA, ESAC) and the ISOC for their schedulling efforts, as well as the \emph{INTEGRAL} Users Group for their support in the observations.


\end{document}